\documentclass[12pt]{JHEP}

\font\blackboard=msbm10 at 12pt
\font\blackboards=msbm7
\font\blackboardss=msbm5
\newfam\black
\textfont\black=\blackboard
\scriptfont\black=\blackboards
\scriptscriptfont\black=\blackboardss

\newcommand{\NP}{{\em Nucl.\ Phys.\ }}

\newcommand{\PL}{{\em Phys.\ Lett.\ }}
\newcommand{\PR}{{\em Phys.\ Rev.\ }}

\newcommand{\ba}{\begin{array}}
\newcommand{\ea}{\end{array}}
\newcommand{\be}{\begin{equation}}
\newcommand{\ee}{\end{equation}}
\newcommand{\bea}{\begin{eqnarray}}
\newcommand{\eea}{\end{eqnarray}}
\newcommand{\beas}{\begin{eqnarray*}}
\newcommand{\eeas}{\end{eqnarray*}}

 \def\CO{{\cal O}} 
\def\CA{{\cal A}}   
 \def\CH{{\cal H}}

\def\darr#1{\raise1.5ex\hbox{$\leftrightarrow$}\mkern-16.5mu #1}

\def\half{{\textstyle{1\over2}}} 
\def\roughly#1{\raise.3ex\hbox{$#1$\kern-.75em\lower1ex\hbox{$\sim$}}}

\def\ap#1#2#3{Ann. Phys. {\bf #1} (#2) #3}

\def\ap#1#2#3{Ann.~Phys. {\bf #1} (#2) #3}
\def\IB{\relax\hbox{$\inbar\kern-.3em{\rm B}$}}
\def\IC{\relax\hbox{$\inbar\kern-.3em{\rm C}$}}
\def\ID{\relax\hbox{$\inbar\kern-.3em{\rm D}$}}
\def\IE{\relax\hbox{$\inbar\kern-.3em{\rm E}$}}
\def\IF{\relax\hbox{$\inbar\kern-.3em{\rm F}$}}
\def\IG{\relax\hbox{$\inbar\kern-.3em{\rm G}$}}
\def\IGa{\relax\hbox{${\rm I}\kern-.18em\Gamma$}}
\def\IH{\relax{\rm I\kern-.18em H}}
\def\IK{\relax{\rm I\kern-.18em K}}
\def\IL{\relax{\rm I\kern-.18em L}}
\def\IP{\relax{\rm I\kern-.18em P}}
\def\IR{\relax{\rm I\kern-.18em R}}
\def\IZ{\relax\ifmmode\mathchoice{
\hbox{\cmss Z\kern-.4em Z}}{\hbox{\cmss Z\kern-.4em Z}}
{\lower.9pt\hbox{\cmsss Z\kern-.4em Z}}
{\lower1.2pt\hbox{\cmsss Z\kern-.4em Z}}
\else{\cmss Z\kern-.4em Z}\fi}
\def\II{\relax{\rm I\kern-.18em I}}

\def\ee#1{{\rm erf}\left(#1\right)}

\def\CA{{\cal A}}

\def\CH{{\cal H}}

\def\CO{{\cal O}}

\def\p{\partial}

\def\inbar{\,\vrule height1.5ex width.4pt depth0pt}

\font\cmss=cmss10 \font\cmsss=cmss10 at 7pt

\def\a{{\alpha}}
\def\ap{{\a}^{\prime}}

\def\s{{\sigma}}

\def\lref{\begingroup\obeylines\lr@f}
\def\lr@f#1#2{\gdef#1{\ref#1{#2}}\endgroup\unskip}
\def\s{\sigma}

\def\and{{a^\dagger_n}}

\def\P{\Psi}

\def\math@note#1{\gdef\@eqnlabel{LAB: #1}}
\title{The singular geometry of the sliver}
\author{Gregory Moore\\
{Department of Physics}\\
{Rutgers University}\\
{Piscataway, NJ 08855, U.S.A.}\\
{\tt gmoore@physics.rutgers.edu}}
\author{Washington Taylor\\
{Center for Theoretical Physics} \\
{MIT, Bldg.  6-306} \\
{Cambridge, MA 02139, U.S.A.} \\
{\tt wati@mit.edu}}

\abstract{We consider ``sliver'' states which act as
projection operators in the matter star product of Witten's cubic
string field theory.  These sliver states, which might be associated
with D$p$-branes, are not finite norm states in the matter string
Hilbert space.  We describe the singularities of these states, and
demonstrate that the sliver states are composed of strings having
singular geometric features.  These singularities take a particularly
simple form in the zero slope limit $\alpha' \rightarrow 0$, where the
star algebra factorizes into a product of the algebra of functions on
space-time and the noncommutative star product of fields associated
with higher string modes.    An analogy to the sliver geometry suggests
a natural mechanism for describing closed string states in open string
field theory.}

\preprint{MIT-CTP-3207, RUNHETC-2001-32, hep-th/0111069}
\begin{document}

\baselineskip16pt
\parskip=4pt

\section{Introduction}
\label{sec:intro}

The celebrated Sen conjectures on open string tachyon condensation
\cite{Sen-universality} have led to a resurgence of interest in string
field theory, particularly in its applications to the physics of
D-branes.  Recently, certain solutions of the matter string field
equations, known as sliver states
\cite{Rastelli-Zwiebach,Kostelecky-Potting,rsz-2}, have received much
attention [5-15].
In this paper we make some simple observations
regarding the structure of these sliver states.  We first consider the
``D-instanton sliver'' in the low-energy limit of string field theory.
In this limit the matter string field star algebra factorizes into a
space-time algebra and an algebra associated with massive string
modes.  The space-time part of the D-instanton sliver becomes a
function which acts as a projection operator in an algebra of
functions on space-time. While the appearance of a projection 
operator is expected from a number of viewpoints, 
a novel point is that  this function is {\it
discontinuous}.  This discontinuity can be associated with a
geometrical localization of the state to strings whose midpoints lie
at the origin of space-time---a condition which can be represented
algebraically as the constraint
\begin{equation}
\hat{x}_0 | \widetilde{\Xi}_0 \rangle = 0,
\label{eq:c1}
\end{equation}
where $\hat{x}_0$ is the operator measuring the center of mass of the
string and $| \widetilde{\Xi}_0 \rangle$ is the $\alpha' \rightarrow
0$ limit of the D-instanton sliver state $| \Xi_0 \rangle$.  This
discontinuity takes the state $| \widetilde{\Xi}_0 \rangle$ outside
the string Hilbert space, as this state formally has vanishing norm.
A similar type of discontinuity occurs in the full D-instanton sliver
state \footnote{The existence of a singularity in the sliver state was
first determined by Rastelli, Sen, and Zwiebach
\cite{rsz-private}}.  We show that in this case the discontinuity can
be associated with the algebraic condition
\begin{equation}
\hat{x}(\pi/2) \; | {\Xi}_0 \rangle = 0,
\label{eq:c2}
\end{equation}
where $\hat{x} (\pi/2)$ is the position of the string midpoint.  This
condition has the geometrical meaning that the functional on the space
of string configurations associated with $| \Xi_0 \rangle$ has support
only on strings whose midpoint is fixed to live at the origin ({\it
i.e.}, on the locus of points defining the D-instanton).  When the
sliver state is taken to be extended in some space-time dimensions
(such as the original D25-brane sliver), a related but distinct
singular structure arises in the directions along which the associated
brane is extended.  We find that the geometrical significance of this
singularity is that the strings composing the higher-dimensional
sliver states can essentially ``split'' into two pieces at the
midpoint, giving a pair of independent strings on the right and left
halves of the string world-sheet.  Algebraically, this condition is
given by
\begin{equation}
\frac{\delta}{ \delta  x_{\lambda} (\sigma)}
\langle x (\sigma)  | {\Xi} \rangle = 0,
\label{eq:c3}
\end{equation}
where $x_\lambda = 
\lambda \theta (\sigma -\pi/2) $ is a step function with a jump
discontinuity at the string midpoint.
This result suggests that the sliver states may actually be related to
a pair of D$p$-branes.  Recent numerical evidence also may support this
possibility \cite{Hata-Moriyama}.

The observations we make here are of relevance in developing further
the analogies between string field algebras and operator algebras such
as $C^*$ algebras, and for understanding the relationship between the
the vacuum string field theory proposed by Rastelli, Sen and Zwiebach,
and Witten's cubic string field theory.  Furthermore, a similar
mechanism to that discussed here for the sliver states can give rise
to singular squeezed open string states which have periodic boundary
conditions on the open string, and which thus represent closed string
states.

In Section 2, we review the 3-string vertex of Witten's open bosonic
string field theory and the sliver states.  In Section 3, we consider
the zero-slope limit of the sliver, and discuss the singular nature of
this state.  In Section 4, we describe the singular features of the
D-instanton and higher-dimensional D$p$-brane slivers.  In Section 5
we propose a class of states with similar singular structure to the
sliver, which we believe represent closed string states.  Section 6
contains conclusions.

Related work on the singular structure of the sliver has been done by
Rastelli, Sen, and Zwiebach, and will appear in a publication by those
authors \cite{rsz-forthcoming}.

\section{D-Branes and the Sliver in String Field Theory}
\label{sec:background}

\subsection{String field theory and the star product}
\label{sec:SFT}

In this subsection we briefly review the star product of matter fields
in Witten's cubic string field theory \cite{Witten-SFT} and fix
notation.    We primarily follow here the conventions of \cite{rsz-2},
so when comparing with
\cite{Gross-Jevicki-1,Gross-Taylor-I,Gross-Taylor-II} factors of $p (x)$ should be
multiplied (divided) by $\sqrt{2}$.

In this paper we are concerned with the matter sector of string field
theory.  The matter fields $x^\mu$ on the
string,  $(\mu \in\{1, \ldots, 26\})$, can be decomposed in 
Fourier modes through
\begin{equation}
x(\s) =  x_0+ \sqrt{2}\sum_{n=1}^\infty{x_n\cos(n\s)}\,, \;
0 \leq \sigma \leq \pi\,.
\quad
\label{eq:x}
\end{equation}
(We will drop most spatial indices in this paper for clarity.)
The modes in (\ref{eq:x})  can be related to creation and annihilation
operators $a^{\dagger}_n, a_n$ satisfying $[a_n, a^{\dagger}_m] =
\delta_{n, m}$
through
\begin{eqnarray}
\hat x_n  =  {i \over  {\sqrt{2n}} }
\left(a_n-a^\dagger_n\right) & \hspace{0.5in} &
         \hat   p_n =-i
{\p \over \p x_n} =  \sqrt{ n\over 2 } \left(a_n+a^\dagger_n\right)
         \label{eq:xp}
\end{eqnarray}
for $n \neq 0$, and through
\begin{eqnarray}
\hat x_0  =  {i \over  {{\sqrt{2}}} }
\left(a_0-a^\dagger_0\right) & \hspace{0.5in} &
            \hat p_0 =-i
{\p \over \p x_0} =   \frac{1}{ \sqrt{2}}   \left(a_0+a^\dagger_0\right)
         \label{eq:xp0}
\end{eqnarray}
for the zero modes.

A matter string field $\Psi$ is a state in the string Hilbert space
${\cal H}$, a basis for which is given by the set of states produced
by acting with a finite number of matter oscillators $a^{\dagger}_n, n
\geq 0$, on the matter vacuum $| 0 \rangle$ annihilated by $a_n$ for
all $n \geq 0$.  It is often convenient to describe matter string
fields in terms of a momentum basis of states
\begin{equation}
| p \rangle = \frac{1}{ ( \pi)^{1/4}} 
\exp \left[-\frac{1}{2}p^2
+ \sqrt{2}a_0^{\dagger} p -\frac{1}{2}(a^{\dagger}_0)^2 \right]
| 0 \rangle
\label{eq:p-basis}
\end{equation}
satisfying $\hat p_0 | k \rangle = k | k \rangle$.  The matter string field
$\Psi$ can be thought of as a functional $\Psi[x (\sigma)]$.  For
well-behaved states in the Fock space (such as those given by acting
on the vacuum with a finite number of raising operators
$a^{\dagger}_n$), this functional corresponds to a well-behaved
function on the countable set of string modes $\{x_n\}$.

The star product of matter string fields corresponds to a map
$\star:{\cal H}\otimes{\cal H}\rightarrow{\cal H}$.  For well-behaved
Fock space states this product is associative.  The star product is
schematically defined by ``gluing'' the right half of one string to
the left half of another with a delta function interaction through
\begin{equation}
         \left(\P \star   \Phi\right) [z(\s)]
         \equiv
\int
\prod_{{0} \leq \s \leq {\pi\over 2}} dy(\s) \; dx (\pi -\sigma)
\prod_{{\pi\over 2} \leq
\s \leq \pi}
\delta[x(\s)-y(\pi-\s)]
\;   \P [x(\s)]  \Phi [y(\s)]
\label{eq:matter-star}
\end{equation}
\begin{eqnarray}
z(\s) & = &x(\s) \quad {\rm for} \quad {0} \leq \s \leq {\pi\over 2}\, ,
\nonumber\\
z(\s) & = &y(\s)\quad {\rm for} \quad   {\pi\over 2} \leq \s \leq \pi\, .
\nonumber
\end{eqnarray}
This definition can be made more precise using a mode decomposition of
the string.   An explicit calculation of the three-string vertex $|
V_3 \rangle \in{\cal H} \otimes{\cal H} \otimes{\cal H}$ satisfying
\begin{equation}
| \Psi \star \Phi \rangle =
(\cdot \otimes \langle \Psi | \otimes \langle \Phi |)\; | V_3 \rangle
\end{equation}
in terms of the string modes was given in
\cite{Gross-Jevicki-1,cst,Samuel,Ohta}.  In the matter sector, this
three-string vertex is given by
\begin{eqnarray}
| V_3 \rangle  & = & \int dp^{(1)}\, dp^{(2)}\, dp^{(3)}\, \;
\delta^{26} (p^{(1)} + p^{(2)} + p^{(3)}) \label{eq:vertex-p}
\\
& &\hspace{0.3in}
\exp \left(-\frac{1}{2}\sum_{r, s \leq 3}
\left[  \sum_{m, n \geq 1} V^{rs}_{m n} (a^{(r)\dagger}_m \cdot
a^{(s) \dagger}_n)
+2  \sum_{m \geq 1} V^{rs}_{m0} (a^{(r)\dagger}_m \cdot
p^{(s)}) + V^{rs}_{00} (p^{(r)} \cdot p^{(s)})
\right] \right)\nonumber\\
& &\hspace{0.6in}
\left( | p^{(1)} \rangle \otimes | p^{(2)} \rangle \otimes
| p^{(3)} \rangle \right) \nonumber\,
\end{eqnarray}
where the Neumann coefficients $V^{rs}_{mn}$ are  computable
constants (see, e.g. \cite{rsz-2}).   Using momentum conservation,
these coefficients can be chosen such that $V^{rs}_{00} = 0$ when $r \neq
s$, and $\sum_{r = 1}^{3}V^{rs}_{0n} = 0$.  These coefficients have
the symmetry $V^{rs}_{mn}= V^{sr}_{nm}$ and cyclic symmetry under $(r,
s) \rightarrow  (r \, ({\rm mod}\, 3) + 1, s\, ({\rm mod}\, 3) + 1)$.

Using (\ref{eq:p-basis}) and performing the 
Gaussian integrals on $p$,  the vertex (\ref{eq:vertex-p}) can be
rewritten as \cite{Gross-Jevicki-1,rsz-2} (using $b = 2$ in the
notation of \cite{rsz-2})
\begin{eqnarray}
| V_3 \rangle  & = & 
\left( \frac{2 \; \pi^{1/4}}{  \sqrt{3} (1 + V_{00})}  \right)^{26}
\exp \left(-\frac{1}{2}\sum_{r, s \leq 3}
\;  \sum_{m, n \geq 0} V'^{rs}_{m n} (a^{(r)\dagger}_m \cdot
a^{(s) \dagger}_n)
 \right)
\left( | 0 \rangle \otimes | 0 \rangle \otimes
| 0  \rangle \right) \label{eq:vertex-0}\,
\end{eqnarray}
where $V_{00} = V^{rr}_{00}$ and
\begin{eqnarray}
V'^{rs}_{mn} & = &  V^{rs}_{mn}
-\frac{ 1}{1 + V_{00}} \sum_{t}V^{rt}_{m0}V^{ts}_{0n}  \nonumber\\
V'^{rs}_{m0} & = &  V'^{sr}_{0m} =
\frac{\sqrt{2}}{1 + V_{00}}
V^{rs}_{m0}   \label{eq:v-relations}\\
V'^{rs}_{00} & = &  
\frac{2}{3 (1 + V_{00})} 
+ \delta^{rs} \left( 1-2/(1 + V_{00}) \right)\,.
\nonumber
\end{eqnarray}
The expression (\ref{eq:vertex-p}) is useful for states of fixed
momentum, while the expression (\ref{eq:vertex-0}) is useful for
states associated with objects localized in space-time (such as
D-instantons).  By using (\ref{eq:p-basis}) in only a subset of $k$
dimensions, the vertex can be written in a form natural for objects of
codimension $k$ in space-time.

\subsection{Matter projectors and the sliver state}
\label{sec:projectors}

The three-string vertex defines an algebra structure on the 
space of string fields. 
Given the close relations between D-branes, noncommutative 
solitons, and K-theory, and the central role of projection 
operators in the latter two subjects, it is natural to search 
for projection operators in the string field algebra. 
Indeed, projection  operators in the matter string field star algebra seem to
be naturally related to single and multiple D-brane configurations in
target space \cite{rsz-3,Gross-Taylor-I,rsz-4}, although the details
of this correspondence in the ghost sector are not yet fully
understood.  

One particularly natural set of projection
operators of the string field star algebra
are those related to the ``sliver state'' found in
\cite{Rastelli-Zwiebach}.  It was shown in
\cite{Kostelecky-Potting,rsz-2} that the matter projection equation
\begin{equation}
\Psi = \Psi \star \Psi
\end{equation}
is satisfied by the zero-momentum sliver state, which is, 
by definition: 
\begin{equation}
| \Xi \rangle =
\left[ \det (1-Z) \det (1 + T) \right]^{13}
\exp \left[ -\frac{1}{2}\sum_{n,m\geq 1} 
 a_m^{\dagger} S_{mn} a_n^{\dagger} \right]
| p = 0 \rangle
\label{eq:sliver}
\end{equation}
where
\begin{equation}
T = \frac{1}{2Z}
\left( 1 + Z-\sqrt{(1 + 3Z) (1-Z)} \right)\,,
\end{equation}
\begin{equation}
Z = CV,
\end{equation}
\begin{equation}
S = CT,
 \label{eq:s}
\end{equation}
$V = V^{rr}$ and $C_{mn} = \delta_{mn} (-1)^m$ are all infinite
matrices indexed by $m, n \geq 1$.
The state (\ref{eq:sliver}) is thus defined in terms of the
Neumann coefficients $V^{rs}_{mn}$ appearing in the three-string vertex
(\ref{eq:vertex-p}).  Replacing these coefficients with the
coefficients $V'$ appearing in (\ref{eq:vertex-0}) leads to a
localized  (``D-instanton'') form of the sliver state
\begin{equation}
| \Xi_0 \rangle =
\left( \frac{\sqrt{3}}{2 \pi^{1/4}}  (1 + V_{00}) \right)^{26}
\left[ \det (1-Z') \det (1 + T') \right]^{13}
\exp \left[ -\frac{1}{2} \sum_{n,m\geq 0} 
a_m^{\dagger} S'_{mn} a_n^{\dagger} \right]
| 0 \rangle\,,
\label{eq:sliver-0}
\end{equation}
where $T', Z',$ and $S'$ are defined as above, but using $C'$ and $V'$
with indices $m, n \geq 0$ instead of $C$ and $V$ with $m, n \geq 1$.
If we carry out the replacement of $V$ with $V'$ in a subset of $25-p$
of the spatial dimensions, we arrive at a hybrid sliver given by
(\ref{eq:sliver}) in $p + 1$ space-time dimensions and
(\ref{eq:sliver-0}) in the remaining $25-p$ dimensions.  This sliver
is - hypothetically - associated with a D$p$-brane.

The states $|\Xi \rangle, | \Xi_0 \rangle$ (and their D$p$-brane
relatives) are not finite norm states in the string Fock space, as we
will discuss in Section 4.  We will first find it instructive,
however, to consider a simple limit of these states in which the
massive string modes decouple.

\section{Zero-slope slivers}
\label{sec:main}
 
\subsection{The zero-slope limit}
\label{sec:limit}

In \cite{wittencomment} Witten pointed out 
that in a certain low energy limit 
the string field algebra should, in some sense, factorize
\begin{equation}
\CA \to \CA_0 \otimes \CA_1 
\label{eq:factorization}
\end{equation}
where, for  open strings on $R^{26}$ with Neumann 
boundary conditions there is a well-defined 
subalgebra $\CA_0 $ of $\CA$ defined by the $p=0$ sector
while  ${\cal A}_1$ is
the $C^*$ algebra of functions on space-time.  When $B_{\mu \nu} = 0$
this algebra is commutative, and when $B_{\mu\nu}$ is 
nonzero (and the limit is taken appropriately as in 
\cite{sw,hklm, wittencomment})
 ${\cal A}_1$ is 
related to the noncommutative Moyal algebra of functions on $R^{26}$. 

There are several conceptual issues raised by
trying to assign a precise meaning to (\ref{eq:factorization}). 
We will take the following pragmatic route. 
(See also the paper by Schnabl \cite{schnabl} for related remarks.) 
 
Let us consider the 3-string vertex for Chern-Simons bosonic open
string field theory with Neumann boundary conditions in a family of
closed string backgrounds parametrized by constant fields $G_{\mu\nu},
B_{\mu\nu}$ on $R^{1,25}$.  The string field vertices have been worked
out in \cite{sugino,kawano}.  These expressions are generally written
in terms of oscillators that depend on the metric:
\begin{equation}
[a^\mu_n, a^\nu_m] =  G^{\mu\nu} \delta_{n+m,0}\,.
\end{equation}
In the absence of a $B$ field, the three-string vertex $| V_3 \rangle$
takes the form (\ref{eq:vertex-p}), where contractions are understood
to be done using the metric $G_{\mu \nu}$.
When thinking about the deformation of algebras in terms 
of explicit structure constants it is useful to keep 
the basis fixed and to let   the structure constants alone 
vary. 
For a family of metrics with open string 
metric $G_{\mu\nu} = t^2 G^{(0)}_{\mu\nu}$ 
we define $\beta_n^\mu = t a_n^\mu$, so that 
\begin{equation}
[\beta^\mu_n, \beta^\nu_m] = (G^{(0)})^{\mu\nu} \delta_{n+m,0}
\end{equation}
We construct a basis for the Fock space using the oscillators 
$\beta_n^\mu$, ($n\not=0$); our statements about limits are statements 
about the $t$-dependence of matrix elements in this basis.

There are three kinds of terms in the string field vertex that we must
discuss:  

1. Type I: 
$$
a_n^{(r),\mu} G_{\mu\nu} a^{(s),\nu}_m V^{rs}_{nm} 
=\beta_n^{(r),\mu} G^{(0)}_{\mu\nu} \beta^{(s),\nu}_m V^{rs}_{nm}
\qquad n,m\geq 1 
$$
Since the Neumann coefficients $ V^{rs}_{nm}$ are $t$-independent, 
these terms  clearly lead to $t$-independent matrix elements in 
the Fock space basis of $\beta$'s. 

2. Type II: 
The cross terms between the zero modes and the oscillators 
scale like 
\begin{equation}
\sqrt{\ap}p^{(r)}_\mu   a^{(s),\mu}_n V^{rs}_{0n} 
= t^{-1} \sqrt{\ap} p^{(r)}_\mu   \beta^{(s),\mu}_n V^{rs}_{0n} 
\end{equation}
These go to zero for $t\to \infty$ at fixed $\sqrt{\ap} p_\mu$, and 
even go to zero if we take $p_{\mu}= t^\theta q_{\mu}$
for $\theta<1$, and $q_\mu$ fixed. 

3. Type III: 
\begin{equation}
\alpha' p_{\mu}^{(r)} G^{\mu\nu} 
p_{\nu}^{(s)} V^{rs}_{00} \sim {1\over t^{2-2\theta}} \to 0 
\end{equation}

In particular, if we define $\epsilon = t^{-1+\theta}$ then 
our scaling limit 
has the same effect as having
  the terms $V_{00}^{rs}$ in the vertex scale with 
$\epsilon^2$ for $\epsilon\to 0$ while $V_{0n}$, $n\geq 1$ scale like 
$\epsilon$. (If we include a noncommutativity parameter and 
take an appropriate limit \cite{sw,hklm}  limit then $\theta=1/2$ is preferred to get a nontrivial 
limit for the phase prefactor.)

\subsection{The star product in the zero-slope limit}
\label{sec:star-limit}

Let us now replace
\begin{eqnarray}
V_{mn} & \rightarrow & V_{mn}  \nonumber\\
V_{m0} & \rightarrow & \epsilon V_{m0}  \label{eq:tv}\\
V_{00} & \rightarrow & \epsilon^2 V_{00}\,,  \nonumber
\end{eqnarray}
and take the limit $\epsilon \to 0 $. 
One finds from (\ref{eq:vertex-p}) that the vertex factorizes
\begin{equation}
| V_3 \rangle \rightarrow | V_3^{(0)} \rangle \otimes | V_3^{(1)} \rangle\,.
\label{eq:factorv}
\end{equation}
We will now give explicit formulae for the vertices $| V_3^{(0)}
\rangle$ and $| V_3^{(1)} \rangle$, thereby defining the algebras
${\cal A}_0$ and ${\cal A}_1$ in (\ref{eq:factorization}). 

The $p = 0$ algebra ${\cal A}_0$ is defined through the three-string vertex
\begin{eqnarray}
| V^{(0)}_3 \rangle  & = & 
(\sqrt{2 \pi})^{26}
\exp \left(-\frac{1}{2}\sum_{r, s \leq 3}
\;  \sum_{m, n \geq 1} V^{rs}_{m n} (a^{(r)\dagger}_m \cdot
a^{(s) \dagger}_n)
 \right)\nonumber
\left( | 0 \rangle \otimes | 0 \rangle \otimes
| 0  \rangle \right) \label{eq:vertex-a0}\,.
\end{eqnarray}
The space-time algebra ${\cal A}_1$ is defined through
\begin{equation}
(f \star g) (p^{(3)}) =
\frac{1}{ (2 \pi)^{13}}  \int dp^{(1)}\, dp^{(2)}\,
\delta^{26} (p^{(1)} + p^{(2)} - p^{(3)}) 
f (p^{(1)}) g (p^{(2)})
\label{eq:vertex-a1}
\end{equation}
which is the expression in Fourier space of the pointwise product
\begin{equation}
(f \star_{\rm point} g) (x) := f (x) g (x)\,.
\label{eq:simple-product}
\end{equation}

Let us now consider the representation (\ref{eq:vertex-0}) of the
three-string vertex for states localized in space-time.  In the limit
$\epsilon \rightarrow 0$, the coefficients $V'$ found by inserting
(\ref{eq:tv}) 
in (\ref{eq:v-relations}) go to
\begin{eqnarray}
V'^{rs}_{mn} & \rightarrow & V^{rs}_{mn} +{\cal O} (\epsilon^2)  \nonumber\\
V'^{rs}_{m0} & \rightarrow & 0 +{\cal O} (\epsilon)  \label{eq:new-vp}\\
V'^{rs}_{00} & \rightarrow & \frac{2}{3}- \delta^{rs}
+{\cal O} (\epsilon^2) \,.  \nonumber
\end{eqnarray}
Again, the algebra factorizes.  The space-time independent part of the
algebra, ${\cal A}_0$, is unchanged since $V'_{mn} \rightarrow
V_{mn}$.  The space-time algebra ${\cal A}_1$ is now represented by
the three-string vertex 
\begin{eqnarray}
| V^{(1)}_3 \rangle  &= &
\exp \left[ \frac{1}{6} \left[
 (a_{(1)}^{\dagger})^2+ (a_{(2)}^{\dagger})^2
+ (a_{(3)}^{\dagger})^2\right]
-\frac{2}{3}  \left[
 a_{(1)}^{\dagger} a_{(2)}^{\dagger}
+ a_{(2)}^{\dagger} a_{(3)}^{\dagger}
+ a_{(3)}^{\dagger} a_{(1)}^{\dagger} \right] \right]
\\
& &\hspace{0.3in}
\times\left( \frac{ \sqrt{2}}{  \sqrt{3} \; \pi^{1/4}}  \right)^{26}
\left( | 0 \rangle \otimes | 0 \rangle \otimes
| 0  \rangle \right)\nonumber
\end{eqnarray}
This vertex encodes the same $C^*$ algebra as
(\ref{eq:simple-product}). This may be proved by 
straightforward manipulation of harmonic oscillators 
as follows. The generating function for Hermite 
functions is 
\begin{equation}
\langle x \vert e^{\sqrt{2} t a^\dagger} \vert 0 \rangle  = 
\sum {(\sqrt{2} t)^n \over \sqrt{n!}} \psi_n(x)  
 = \pi^{-1/4} e^{-\half x^2 + t^2 - 2i tx} 
\end{equation}
We may multiply these pointwise in $x$ and re-express the 
result in terms of harmonic oscillators.
The result is 
\begin{equation}
e^{\sqrt{2} t_1 a^\dagger} \vert 0 \rangle *_{\rm point} 
 e^{\sqrt{2} t_2 a^\dagger} \vert 0 \rangle = 
\pi^{-1/4}\sqrt{2\over 3} 
e^{{1\over 3} (t_1^2+t_2^2)-{4\over 3} t_1 t_2} e^{-{2\sqrt{2}\over 3}(t_1+t_2)
 a^\dagger + {1\over 6} (a^\dagger)^2 } 
\vert 0 \rangle.
\label{eq:pointosc}
\end{equation}
The right hand side of (\ref{eq:pointosc}) can also be written as 
\begin{equation}
{}_1\langle 0 \vert e^{\sqrt{2} t_1 a_1}
{}_2\langle 0 \vert e^{\sqrt{2} t_2 a_2}\vert V\rangle
\end{equation}
where 
\begin{equation}
\vert V \rangle = \pi^{-1/4}\sqrt{2\over 3} 
\exp\Biggl[{1\over 6}[ (a_1^\dagger)^2+(a_2^\dagger)^2+(a_3^\dagger)^2]
-{2\over 3} [a_1^\dagger a_2^\dagger +  
a_1^\dagger a_3^\dagger +a_2^\dagger a_3^\dagger ]\Biggr]\vert 0 \rangle_1 
\otimes\vert 0 \rangle_2 \otimes \vert 0 \rangle_3. 
\end{equation}

\subsection{The limit of the space-time filling sliver}
\label{sec:25-limit}

The projector associated with a space-filling D25-brane is given by
the zero-momentum matter sliver state (\ref{eq:sliver}).  Since the
indices $m, n$ in (\ref{eq:sliver}) range from 1 to infinity, this
state is space-time independent, and in the limit $\epsilon \rightarrow 0$
can be thought of as a product of a nontrivial state in ${\cal A}_0$,
given again by (\ref{eq:sliver}), and the function $f(x)=1$
({\it i.e.} the state $\sqrt{2 \pi} |
p_0 = 0 \rangle$) in ${\cal A}_1$, which clearly acts as a projection
operator under (\ref{eq:simple-product}).

\subsection{The limit of the lower-dimensional slivers}
\label{sec:skyscraper}

In the limit $\epsilon \rightarrow 0$, we see from (\ref{eq:v-relations})
that 
\begin{eqnarray}
V'_{00} \rightarrow -1/3, & \hspace*{1in} &  V'_{0n}, V'_{n0} \rightarrow 0\\
Z'_{00} \rightarrow -1/3, & \hspace*{1in} &   Z'_{0n}, Z'_{n0} \rightarrow 0\\
T'_{00} \rightarrow -1, & \hspace*{1in} &   T'_{0n}, T'_{n0} \rightarrow 0\\
S'_{00} \rightarrow -1, & \hspace*{1in} &   S'_{0n}, S'_{n0} \rightarrow 0\,.
\end{eqnarray}
%
%
The advantage of the low energy limit now becomes apparent.  It
separates out the space-time dependence of the sliver states
associated with lower-dimensional D$p$-branes, and makes manifest the
existence of an eigenvector of $S'$ and of $T'$ of eigenvalue $-1$.
This eigenvector is associated with the zero-mode raising operator
$a_0^{\dagger}$, and gives the sliver a singular structure which we
now discuss.

Separating out the space-time dependence encoded in $S'_{00}$, etc.,
we have
\begin{equation}
| \widetilde{\Xi}_0 \rangle =
\lim_{t \rightarrow 0} | \Xi_0 \rangle = A_1 \otimes A_0
\label{eq:D-instanton-limit}
\end{equation}
where $A_0 = | \Xi \rangle$ and $A_1$ is given by 
\begin{eqnarray}
A_1 & = &  \lim_{s \rightarrow -1} 
\left[  \pi^{1/4} \sqrt{1 + s} \right]^{26}
 \exp \left(-\frac{s}{2} (a_0^{\dagger})^2\right) | 0 \rangle.
\end{eqnarray}
The overall  normalization of $A_1$ is determined 
by normalizing $A_0$ so that it becomes a projector 
with respect to $V^{(0)}$. Since the product is a projector, 
the normalization of $A_0$ fixes $A_1$ to be a projector. 

We can translate from the squeezed state representation to a
function of $x_0$ using (for each coordinate direction)
\begin{equation}
\langle x \vert \exp \left(-\frac{1}{2}s (a_0^{\dagger})^2\right)| 0 \rangle
= \pi^{-1/4} \frac{1}{\sqrt{1 + s}} 
\exp \left(-\frac{\lambda}{ 2}x^2 \right)
 \label{eq:function-squeezed}
\end{equation}
where
\begin{equation}
\lambda = \frac{1-s}{1 + s} \,.
\end{equation}
Thus, we have
\begin{eqnarray}
A_1(x_0) & = &  \lim_{\lambda \rightarrow \infty} 
 \exp \left(-\frac{\lambda}{ 2}x_0^2 \right)  \nonumber\\
& = & \left\{
\begin{array}{l}
0, \;\;\;\;\; x_0 \neq 0,\\
1, \;\;\;\;\; x_0 = 0
\end{array}
\right. \label{eq:unit-d}
\end{eqnarray}

The function (\ref{eq:unit-d})  is indeed a projection operator under 
the pointwise product
(\ref{eq:simple-product}) in the sense that $A_1^2=A_1$. The 
surprising point here is that the limit is not a continuous 
projector. 
The singular nature of this sliver state can be expressed
algebraically by the simple condition that
\begin{equation}
x^\mu_0 \; | \widetilde{\Xi}_0 \rangle = 0\,
\label{eq:sliver-condition-1}
\end{equation}
in the directions transverse to the sliver.  This condition states
geometrically that the strings of which the zero-slope limit of the
sliver are comprised have a center of mass which is confined to the
$(p + 1)-$dimensional hypersurface of the associated D$p$-brane.

\section{Singular geometry of D$p$-brane slivers}
\label{sec:geometry}

In this section we show that, even without taking the zero-slope limit,
the sliver states associated with D$p$-branes have singularities which
can be interpreted geometrically.

\subsection{Formal eigenvalues of $S$ and $S'$}
\label{sec:bad}

The sliver states $| \Xi \rangle$ and $| \Xi_0 \rangle$ are described
as squeezed states in (\ref{eq:sliver}, \ref{eq:sliver-0}).  These
squeezed states fail to have finite norm if the associated matrices
$S, S'$ have eigenvalues of $\pm 1$.  It was found numerically by Rastelli,
Sen, and Zwiebach that such eigenvalues indeed seem to occur
\cite{rsz-private}.  In this subsection we give a simple analytic
derivation of the eigenvectors associated with these problematic
eigenvalues; another approach to determining these eigenvectors will
be discussed in \cite{rsz-forthcoming}.

Let us begin with the D-instanton sliver $| \Xi_0 \rangle$.  It is
convenient to recall some notation from
\cite{Gross-Jevicki-1,Gross-Taylor-I}.  Note that in those references
slightly different notation is used---in particular, the matrices we
refer to as $U', V'$ here are referred to as $U, V$ in those
references.

The D-instanton sliver is constructed from the matrix $V'_{mn}$ of Neumann
coefficients with indices $m, n \geq 0$.  It was shown in
\cite{Gross-Jevicki-1} that this matrix can be written as
\begin{equation}
V' =\frac{1}{3} \left( C' + U' + \bar{U}' \right)
\label{eq:v}
\end{equation}
in terms of a matrix $U$ implicitly defined through the overlap
conditions
\begin{eqnarray}
(1-Y') E' (1 + U') & = &  0\label{eq:yu}\\
(1+Y') E'^{-1} (1 - U') & = &  0\,. \nonumber
\end{eqnarray}
These conditions are given in terms of the matrices \footnote{Note
that this matrix, like equation (\ref{eq:xp0}), differs from the
notation of \cite{Gross-Jevicki-1,Gross-Taylor-I} by a factor of
$\sqrt{2}$ in the 0-index component.}
\begin{equation}
E'^{-1}_{mn} = \delta_{mn} (\sqrt{n} + \delta_{m0})
\end{equation}
and
\begin{equation}
Y' = -\frac{1}{2}C' - \frac{\sqrt{3}}{2}  i C' X'\,,
\label{eq:y}
\end{equation}
which is in turn defined through the matrix $X'$ with nonzero matrix
elements
\begin{eqnarray}
X'_{2k+1,2n}=X'_{2n,2k+1} & = & { 4(-1)^{k+n}(2k+1)\over
\pi\left({(2k+1)^2-4n^2}\right)} \;\;\;  \quad  n \geq 1 \, , \label{eq:X}\\
\quad X'_{ 2k+1,0} =X'_{0,2k+1} & = & 
{  2 \sqrt{2}(-1)^{k}\over \pi{(2k+1)}}\, .\nonumber
\end{eqnarray}

It is useful to understand these matrices more conceptually. 
Note that the Hilbert space $L^2[0,\half \pi]$ has two 
distinct orthonormal bases: 
\begin{equation}
\varphi_0 := \sqrt{2\over \pi} \qquad \varphi_n := \sqrt{4\over \pi} \cos(2n\sigma) \qquad n=1,\dots
\label{eq:evens} 
\end{equation} 
and 
\begin{equation} 
\psi_k := \sqrt{4\over \pi} \cos[(2k+1)\sigma] \qquad k=0,1,\dots 
\label{eq:odds}
\end{equation} 
The matrix $X'$ is simply the unitary transformation between these bases: 
\begin{eqnarray}
\psi_k & = & \sum_{n=0}^\infty X'_{2k+1,2n} \varphi_n \label{eq:unitary} \\ 
\varphi_n & = & \sum_{k=0}^\infty X'_{2n,2k+1} \psi_k \nonumber
\end{eqnarray}

\def\CH{{\cal H}} 
On the other hand, the Hilbert space $\CH = L^2[0,\pi]$ has a 
natural decomposition into $\pm 1$ eigenspaces of the 
flip operator $f(\sigma) \to f(\pi-\sigma)$, 
namely  $\CH = \CH^+ \oplus \CH^-$. 
We refer to these as the even and odd sectors. Now, ${1\over \sqrt{2}} \varphi_n$
and ${1\over \sqrt{2}} \psi_k$  extend to 
form orthonormal bases for $\CH^\pm $, respectively.  Therefore, the 
unitary transformation (\ref{eq:unitary}) defines a unitary 
isomorphism $\CH^+ \to \CH^-$. This map is just 
the map 
\begin{equation}
f\to f \chi_{[0,\half \pi]} - f\chi_{[\half \pi, \pi]},
\label{eq:left-right}
\end{equation}
where $\chi_A$ is the characteristic function of a measureable 
set $A$. Indeed   the main utility of $X'$ in 
\cite{Gross-Taylor-I}, was that it related 
 modes on the full string to modes on the left
and right halves of the string, and it was used extensively  
 to develop a ``split string'' formalism for the
matter sector of string field theory. 

Now, formally, the map (\ref{eq:left-right}) annihilates the 
$\delta$-function $f(\sigma) = \delta(\sigma - \half \pi)$, so 
we might expect $X'$ to have a null eigenvector. Indeed, one 
can check that 
\begin{equation} 
\sum_{n} X'_{mn} w'_n= 0
\label{eq:nullvector}
\end{equation}
where $w'_n$ is 
given by
\begin{equation}
w'_n = \left\{\begin{array}{ll}
 \frac{1}{ \sqrt{2}}, & n = 0\\
(-1)^{n/2}, & n \; {\rm even}, \;n > 0\\
0, & n \; {\rm odd}
\end{array} \right.
\end{equation}
We stress that the sum in (\ref{eq:nullvector}) is convergent. 
While it might be alarming to find a zero eigenvalue of a 
unitary operator, there is really no mathematical contradiction: 
$w_n'$ is not normalizable, indeed $f(\sigma) = \delta(\sigma - \half \pi)$
is not an $L^2$-function, but rather a tempered distribution. 
Nevertheless, we will now argue that this formal eigenvector plays an 
important role in understanding the meaning of the sliver state.

It is convenient to decompose the various matrices into two by two
block form associated with even and odd sectors, so for example
\begin{equation}
X' =
\left(\begin{array}{cc}
0&  X'_{oe}\\
X'_{eo}& 0
\end{array} \right)\, ,
\end{equation}
and
\begin{equation}
U'=
\left(\begin{array}{cc}
U'_{oo}& U'_{oe}\\
U'_{eo}& U'_{ee}
\end{array} \right),\;\;\;\;\;
\bar{U}'=
\left(\begin{array}{cc}
U'_{oo}& -U'_{oe}\\
-U'_{eo}& U'_{ee}
\end{array} \right).
 \label{eq:u1}
\end{equation}
In terms of these blocks, the even-even sector of the first equation
in
(\ref{eq:yu}) gives
\begin{equation}
E'\, (1 + U'_{ee}) = \frac{-i}{\sqrt{3}}  X'_{eo} \,E'\, U'_{oe}\,.
\label{eq:eetooh}
\end{equation}
Since the RHS vanishes when we act on the left with $w'$, it follows
immediately that the vector $\nu'= E' w'$ is an eigenvector of
the symmetric matrix $U'_{ee}$ with eigenvalue $-1$.
{}From (\ref{eq:v}) this implies that $\nu'$ is an eigenvector of $V'$
with eigenvalue $-1/3$, and hence an eigenvector of $S'$ of eigenvalue
$-1$.  
This conclusion is also supported by a numerical analysis of the
spectrum of $V'$ in level truncation.  Thus, we have identified an
eigenvector $\nu'$ of $S'$ with components
\begin{equation}
\nu'_n  = \left\{\begin{array}{ll}
 \frac{1}{\sqrt{2}}, & n = 0\\
\frac{(-1)^{n/2}}{\sqrt{n}},\; \; & n \; {\rm even}, \;n > 0\\
0, & n \; {\rm odd}
\end{array} \right.
\label{eq:nup}
\end{equation}

The existence of this
(nonnormalizable) eigenvector of eigenvalue $-1$ indicates that 
 the norm of the
D-instanton sliver state vanishes.  We will discuss the geometrical
meaning of this vanishing norm in the following subsection.
 

It is interesting to consider the behavior of the eigenvector $ \nu'$
in the limit $t \rightarrow \infty$ considered in the previous
section.  Changing the metric to $G_{\mu \nu} = t^2 G^{(0)}_{\mu
\nu}$,  we see that the essential change in the relationship
between the modes $x_n$ and the raising and lowering operators $a_n,
a^{\dagger}_n$ is that instead of (\ref{eq:xp}) we have for  $n > 0$
\begin{equation}
\hat{x}^n = \frac{i}{t^2  \sqrt{2}}  (a_n-a_n^{\dagger}).
\label{eq:new-xp}
\end{equation}
Note that the relation (\ref{eq:xp0}) is unchanged as we are keeping
the definitions of $x_0, p_0, a_0$ fixed.
This change means that in the first overlap equation of (\ref{eq:yu}),
the matrix $E'$ must be replaced by the matrix
\begin{equation}
E^{(t)}_{mn} = 
\left\{\begin{array}{ll}
\delta_{mn} \frac{1}{t^2 \sqrt{n}}, \;\;\;\;\; & m > 0\\
\delta_{0n}, & m = 0
\end{array}
\right.
\end{equation}
This change must also be made in the matrix $E'$ appearing on the left
hand side of (\ref{eq:eetooh}).  This means that for an
arbitrary value of $t$, the vector $\nu^{(t)} = E^{(t)}w'$ is an
eigenvector of  $U^{(t)}_{ee}$ with eigenvalue -1.  It is clear that
in the limit this vector approaches
\begin{equation}
\lim_{t \rightarrow \infty} \nu^{(t)}_n = \frac{1}{ \sqrt{2}}  \delta_{n0}\,.
\end{equation}
This is precisely the eigenvector we identified in Section 3 in the $t
\rightarrow \infty$ limit.

Now let us turn to the D25-brane sliver $| \Xi \rangle$.  The matrix
$S$ appearing in the squeezed state description of this sliver
(\ref{eq:sliver-0}) is related to the matrix $V_{mn} (m, n > 0)$ of Neumann
coefficients in the same way that $S'$ is related to $V'$.  Just as
$V'$ can be related to the matrix $U'$ satisfying the overlap
equations (\ref{eq:yu}), the matrix $V$ can be related to a matrix $U$
satisfying similar overlap equations through
\begin{equation}
V =\frac{1}{3} \left( C + U + \bar{U} \right)
\end{equation}
Just as $U'$ is related to the
matrix $X'$ given by (\ref{eq:X}), $U$ is related to the matrix $X$
which is given by simply restricting $X'$ to nonzero indices ($X_{mn}
= X'_{mn}, m, n > 0$).
The details of this
relationship were worked out by Moeller in \cite{Moeller-sliver}. 
It follows immediately from the overlap equations for $U$ that
\begin{equation}
 1-U_{oo} = \frac{i}{ \sqrt{3}} U_{oe}
E^{-1} X_{eo} E\,.
\label{eq:Moeller-equation}
\end{equation}
(This is equivalent to equation (2.31) in \cite{Moeller-sliver}).

Just as $X_{oe}'$ has a null vector $w'$, the matrix $X_{eo}$ has the
null vector $w$ with components $w_{2k + 1} = (-1)^k/(2k + 1)$.  This
null vector can easily be understood geometrically, as these are the
coefficients in an expansion of the constant function on the half
string in terms of odd cosines.  Thus, $X_{eo}' w$ is naturally a
vector whose only nonzero component is in the 0-index direction, so
that it follows immediately that $X_{eo} w = 0$.  From this null
vector and the equation (\ref{eq:Moeller-equation}) it follows that
$\nu = E^{-1} w$ is an eigenvector of $U_{oo}$ with eigenvalue + 1,
and hence an eigenvector of $V$ with eigenvalue +1/3.  In turn, $\nu$
must be an eigenvector of $S$ with eigenvector + 1.  Thus, we have
identified an eigenvector $\nu$ of S with components
\begin{equation}
\nu_n  = \left\{\begin{array}{ll}
\frac{(-1)^{(n-1)/2} }{\sqrt{n}},\; \; & n \; {\rm odd},\\
0, & n \; {\rm even}
\end{array} \right.
\label{eq:nu}
\end{equation}

\subsection{Geometrical interpretation of sliver singularities}

Let us now turn to a discussion of the geometrical significance of the
singular eigenvectors $\nu', \nu$ we found for the matrices $S', S$ in
Section 4.1.  We begin by recalling that the zero-slope limit of the
sliver $| \widetilde{\Xi}\rangle$ is described by a squeezed state 
proportional to
\begin{equation}
e^{-\frac{ s}{2}(a_0^{\dagger})^2} | 0 \rangle\,,
\label{eq:simple-squeezed}
\end{equation}
with $s \rightarrow -1$.  Acting on this state with $a_0$ gives
\begin{equation}
a_0 e^{-\frac{ s}{2}(a_0^{\dagger})^2} | 0 \rangle
= -s a_0^{\dagger} e^{-\frac{ s}{2}(a_0^{\dagger})^2} | 0 \rangle
\end{equation}
so that when $s \rightarrow -1$ the state (\ref{eq:simple-squeezed})
is annihilated by the operator
\begin{equation}
\hat{x}_0 = i (a_0-a_0^{\dagger})/\sqrt{2}\,,
\end{equation}
as stated in (\ref{eq:sliver-condition-1}).

Similar relations hold for the D$p$-brane sliver states, following
from the existence of the eigenvectors $\nu'$ and $\nu$.  First
consider the D-instanton sliver.  
We have in general for any $\sigma$
\begin{equation} \hat x(\sigma) \vert \Xi_0 \rangle = -{i\over \sqrt{2}} 
\biggl( \sum_{n=0}^\infty \xi_n(\sigma) a_n^\dagger \biggr)
\vert \Xi_0 \rangle
\end{equation} 
where 
\begin{equation} 
\xi_n(\sigma) = \zeta_n(\sigma) + \sum_{m=0}^\infty 
\zeta_m(\sigma) S'_{mn} 
\label{eq:zeta}
\end{equation} 
and $\zeta_0(\sigma) = 1 $, $\zeta_m(\sigma) =
\sqrt{2\over m} \cos(m\sigma)$.    
Since $\nu'$ is an eigenvector of
$S'$ with eigenvalue -1, and $\nu' = \frac{1}{\sqrt{2}} \zeta (\pi/2)$,
it follows that
%
\begin{equation}
\hat{x} (\pi/2) | \Xi_0 \rangle =
\sqrt{2} 
\left(\nu'_0 \hat{x}_0 + \sum_{n > 0} \nu'_n \sqrt{n} \hat{x}_n\right)
| \Xi_0 \rangle 
= 0\,.
 \label{eq:D-instanton-condition}
\end{equation}
Similar to the geometric condition on the zero-slope limit of the
sliver, this condition states that the strings comprising the
D-instanton sliver have a midpoint which is constrained to live at the
origin of space-time.  For a general D$p$-brane sliver, this condition
holds in the transverse dimensions, so that the string midpoints are
constrained to live on the D$p$-brane hypersurface.

Now let us consider the D25-brane sliver.  We have in this case
\begin{equation}
\sum_{n} \nu_n a_n | \Xi \rangle
= -\sum_{n, m}\nu_n S_{nm} a^{\dagger}_m | \Xi \rangle
= -\sum_{n} \nu_n a_n^{\dagger} | \Xi \rangle\,.
\end{equation}
This leads to the condition
\begin{eqnarray}
0 & = &   \frac{i}{ \sqrt{2}} 
\sum_{n}\nu_n (a_n + a_n^{\dagger}) | \Xi \rangle \nonumber\\
 & = &  \sum_{k = 0}^{ \infty} 
\left(\frac{(-1)^k}{2k + 1} \right)  \frac{\partial}{ \partial x_{2k + 1}}  
\langle x (\sigma)| \Xi \rangle\,,\label{eq:sliver-flat-direction}
\end{eqnarray}
where the derivatives are interpreted as acting on the state $\langle
x (\sigma)| \Xi
\rangle$ represented as a function of the string modes $\{x_n\}$.
Just as the condition $0 =p_0 | p = 0 \rangle$ indicates that the
string functional associated with the zero momentum ground state is
flat under translations by a constant function $x (\sigma) = \lambda$,
the condition (\ref{eq:sliver-flat-direction}) indicates that the
D25-brane sliver state is described by a string functional invariant
under translations by a function of the form
\begin{equation}
\delta x (\sigma) = \lambda \sum_{k = 0}^{ \infty} 
\frac{(-1)^k}{2k + 1}  \cos (2k + 1) \sigma
= \left\{\begin{array}{ll}
+ \lambda \pi/2,\;\;\;\;\; & 0 < \sigma < \pi/2\\
- \lambda \pi/2,\;\;\;\;\; & \pi/2 < \sigma < \pi
\end{array}
 \right.
\end{equation}
This indicates that the strings comprising the D25-brane sliver can
``split'' into separate right and left half-strings which have
different boundary conditions at the point $\sigma = \pi/2$.  

We have found in this section that the D$p$-brane sliver states are
comprised of strings whose midpoints are bound to the $(p +
1)$-dimensional hypersurface associated with the relevant D$p$-brane,
but that these strings also seem to split naturally into two
independent parts whose endpoints originally associated with the
string midpoint can live at arbitrary points on the relevant
D$p$-brane world-volume.  It is not clear what the physical
interpretation of this geometrical condition should be, but it seems
likely that this picture will play a part in clarifying the role of
D$p$-brane matter sliver projection operators in Witten's string field
theory and the simplified RSZ vacuum string field theory proposal.
{}From the splitting of the strings involved in the D25-brane sliver
state, it is tempting to postulate that this state may be closely
related to a state in string field theory with {\it two} D25-branes
rather than the original one of the perturbative $U(1)$ vacuum of
Witten's cubic string field theory.  Such a two-brane state should
exist in the $U(1)$ cubic string field theory, although as yet there
is no convincing evidence for this state, either numerical or
analytic \footnote{As this paper was being completed,  an interesting
paper appeared by Hata and Moriyama \cite{Hata-Moriyama}, giving
numerical evidence that indeed the sliver state should be associated
with the two D25-brane states based on the energy of the solution of
VSFT proposed in \cite{Hata-Kawano}}.  
Note that it is the fluctuations around a two-brane state, rather
than the two-brane state itself, in which the strings should naturally
split.  Nonetheless, this geometric picture of the sliver as a
condensate of strings connected to a pair of D$p$-branes is very
suggestive.  In the context of the superstring theory, we might expect
a similar picture giving a brane-antibrane pair.

The geometrical picture we have found of the sliver singularities ties
in neatly with previous calculations
\cite{rsz-2,Gross-Taylor-I,Moeller-sliver} which have shown that the
sliver states factorize into a product of Gaussians of the left and
right string oscillators $r_{2k + 1}, l_{2k + 1}$ of the form 
\begin{equation}
\exp
\left( -\frac{1}{2} \;l \cdot M \cdot l-\frac{1}{2} \;r \cdot M \cdot r
\right)\,.
\label{eq:}
\end{equation}
In the case of the D-instanton sliver, this factorization
is natural since each half of the string is associated with a state in
an ND string Hilbert space.  For higher-dimensional branes, each half
of the string is described by a state with NN boundary conditions in
the longitudinal directions.  In both cases the two half strings are
completely decoupled, so it is natural for the full-string state to
take a product form

\subsection{Some mathematical embarrassments}
\label{sec:fussy}

As we have mentioned, the ``eigenvectors'' discussed above are not
normalizable vectors in the Hilbert space $\ell_2$ of square-summable
sequences. Nevertheless, as we have shown, they have very natural
geometrical interpretations. What should we do about this? We do not
really know the answer to this important question, but in this section
we will take the opportunity to point out a few related awkward
mathematical facts which further suggest that subtleties of functional
analysis may be important in understanding this picture.
 
First, the string field product does not take 
Fock space states to well-defined   states  in the 
string field theory Hilbert space. 
Indeed, even the state $\vert 0 \rangle * \vert 0 \rangle $ 
does not actually exist as a rigorously well-defined state in the 
string field theory Fock space. The reason for this is 
that in rigorous treatments of 
Bogoliubov transforms \cite{shale, ruijsenaars} it is shown that 
the  state $\exp[ \half V_{nm} a_n^\dagger a_m^\dagger] \vert 0\rangle $ 
only exists if the operator $V_{nm}$ is Hilbert-Schmidt. 
On the other hand, it is easy to establish the asymptotic 
formula\footnote{Related results appeared in \cite{Romans}.}
\begin{equation} 
V_{n,m}^{1,1} \sim {8\over \pi\sqrt{27} } (-1)^{{n+m\over 2} } 
{(nm)^{1/6} \over n^{4/3} + (nm)^{2/3} + m^{4/3} } (1+ \CO(1/n,1/m)) 
\label{eq:asymptotics}
\end{equation}
for $n,m$ both even, while for $n,m$ both odd we have the same result 
with an overall $-1$ in front. It follows that 
$\sum_{nm}\vert V_{n,m}\vert^2 \sim \int^\infty {dr\over r} = \infty$
and hence it cannot be Hilbert-Schmidt. 
\footnote{Similar allegations can be made against boundary states in 
boundary conformal field theory. In this context the nonexistence 
of the state is not physically important because one always works in practice with 
the state $q^{L_0} \vert B \rangle \rangle$. One can attempt to 
use a similar cure  in string field theory, but this  
 alters the algebra structure in an important way, replacing 
an associative algebra by a homotopy associative $A_{\infty}$ 
algebra. See, for example \cite{nakatsu}. }

It is clearly of interest to know if $S_{nm}$ is Hilbert-Schmidt. 
We can argue that this is not the case by contradiction 
as follows. If $S$ were Hilbert-Schmidt then $T=CS$ would be too. 
Meanwhile,   $T$ satisfies 
\begin{equation} 
Z = CV = T {i \over \sqrt{3}} \biggl( {1\over T- e^{-i \pi/3} } - 
{1\over T - e^{i \pi/3}} \biggr) 
\end{equation}
If $T$ were HS then since it is symmetric  
 it would be self-adjoint, and in particular 
 ${1\over T- e^{\pm  \pi/3} } $ 
would be bounded. Since Hilbert-Schmidt operators form an 
ideal in the algebra of bounded operators, if 
$T$ were Hilbert-Schmidt then $Z= CV $ would be also, 
but we have just seen that this is not the case. 

We should also mention that 
reasoning similar to what we have used above 
can lead to apparent mathematical absurdities. 
Let us consider the identity on matrices: 
\begin{equation} 
X_{oe} E_e^{-4} X_{eo} = E_o^{-4} 
\end{equation} 
This equation only involves 
convergent sums, yet it is rather alarming because 
$E_e^{-4}$ and $E_{o}^{-4}$ are both diagonal 
matrices with disjoint sets of eigenvalues, 
and yet the unitary matrix $X_{eo}$ has 
transformed one into the other! As with our 
null eigenvalue for $X_{eo}$ there is no 
real mathematical contradiction here. The 
unbounded symmetric operator $-{d^2\over d\sigma^2}$ 
acting on smooth functions on $[0,\pi]$ has
 inequivalent, self-adjoint extensions;
the spectral theorem only applies to 
self-adjoint operators.

Finally, one might hope that although expressions in the matter 
theory are ill-defined due to facts such as those we have just 
quoted, nevertheless, the full matter $\otimes $ ghost theory 
will be better behaved. While this might prove to be the case 
the following example should serve as a cautionary tale.  
 Consider a state in $L^2(R^2) = L^2(R)
\otimes L^2(R)$ given by $\psi_a(x,y) = e^{-\half a x^2} e^{-\half
{1\over a} y^2}$.  For all $a$ this state has norm
$\sqrt{\pi}$. Nevertheless, we do not get a Cauchy sequence of states
as $a \to 0$ or $a \to \infty$.  Hence the limiting state $a\to \infty$ 
does not exist as a state in the Hilbert space. 
(it is ``trying'' to approach $\delta(x)\cdot 0$.)

\subsection{Relation to skyscraper sheaves? } 
\label{sec:crazy}

  A useful perspective on the condition
(\ref{eq:D-instanton-condition})  for the D-instanton sliver, and the
limit (\ref{eq:sliver-condition-1}) of this condition as $t
\rightarrow \infty$ can be gained from the split
string formalism \cite{Gross-Taylor-I,Gross-Taylor-II}.  Considering
the string field 
as an operator on the half-string Hilbert space, the space-time
dependence of the string field can be separated out using the string midpoint,
giving a description of the string field as an operator-valued
function on space time
\begin{equation}
\Psi \rightarrow \tilde{\Psi} (\bar{x})\,,
\end{equation}
where $\bar{x}= x (\pi/2)$ is the position of the string midpoint and the
operator $\tilde{\Psi}(\bar{x})$ acts on the usual Hilbert space of
states of an ND string for every value of $\bar{x}$.  
\footnote{The utility of this picture of string field theory was
suggested to us by I.M. Singer.} The operator
resulting from the D-instanton sliver is in this picture an operator which
projects onto the Hilbert space over the point $\bar{x} = 0$.  Thus,
this operator is essentially projecting a general section of an ${\cal
H}_{{\rm ND}}$-bundle over space-time onto a ``skyscraper sheaf'' with a
single ${\cal H}_{\rm ND}$ fiber at the point 0.

There have been several indications over the years that D-branes
localized in spacetime should be formulated in terms of sheaves.
(See, for example, \cite{harvey-moore} and references therein for one
such discussion.) It would be interesting to use string field theory
projectors in combination with the zero-slope limit to derive the
connection between D-branes and coherent sheaves from a more
fundamental starting point.


\section{Closed strings}
\label{sec:closed}

A crucial question on which the considerations of this paper
seem to shed some light is the question of whether and how closed
strings can arise in the language of open string field theory.  We
found that the D25-brane sliver satisfies a nontrivial condition which
essentially splits the string into two parts by allowing an arbitrary
jump discontinuity in the position of the string at the midpoint.  Due
to this condition, the D25-brane matter sliver state has an infinite norm,
analogous to the infinite norm of a state with fixed momentum.  By
choosing an alternative algebraic condition, we can select states
which geometrically behave as closed strings.  Namely, we can
consider a class of squeezed states
\begin{equation}
|\Gamma \rangle =e^{-\frac{1}{2}a_n^{\dagger} K_{nm}a_m^{\dagger}}| 0 \rangle
\end{equation}
built on the usual open string
vacuum which satisfy the algebraic condition
\begin{equation}
\left[\hat{x} (0) -\hat{x} (\pi) \right]
| \Gamma \rangle = 0\,.
\label{eq:closed-relation}
\end{equation}
This condition is satisfied when the vector $\tilde{\nu} = \left[\zeta
(0)-\zeta ( \pi) \right]/2^{3/2}$ (using the notation of (\ref{eq:zeta}))
is an eigenvector of $K$ with
eigenvalue -1.  This 
vector is given by $\tilde{\nu}_n = 1/ \sqrt{n}$ when $n$ is odd, and
$\tilde{\nu}_m = 0$ when $m$ is even.  This vector is  the same as
the vector $\nu$, except for the signs.   

We believe it is possible to construct a large family of states
satisfying (\ref{eq:closed-relation}).  We further 
propose that  these   are precisely
the states given by
writing the closed string Fock space vacuum and excited states  as
functionals of the closed string modes, translating to open string
modes, and rewriting the resulting
states in terms of open string oscillators.  We now present two 
pieces of evidence for this proposal.

For a closed string on the interval $0 \leq \sigma \leq \pi$, we can
perform a mode decomposition analogous to (\ref{eq:x}) through
\begin{equation}
x(\s) =  z_0+ \sqrt{2}\sum_{n=1}^\infty{z_n\cos( 2 n\s)}
+ \sqrt{2}\sum_{m=1}^\infty{y_m\sin( 2 m\s)}\,.
\end{equation}
Just as the matrices $X'$ relate the NN and ND bases of the Hilbert
space of $ L^2$ functions on the interval $[0, \pi/2]$, a closely
related matrix relates the basis $\cos n \sigma$ of functions on the
interval $[0, \pi]$ with NN boundary conditions and the basis $\cos 2n
\sigma, \sin 2m \sigma$ on which one could impose
 periodic boundary conditions.  The
transformation between these bases is given by
\begin{eqnarray}
z_n & = & x_{2n}, \;\;\; \;\;\;\;\;\;\;\;\;\; \;\;\;\;\;\;
\;\; \;\;\;\;\; \;\;\;\;\; n = 1, 2, \ldots\nonumber\\ 
y_m & = & \sum_{k = 0}^{\infty}Y_{2m, 2k + 1} \;x_{2k +
1} \;\;\;\;\; \;\;\;\;\; m = 1, 2, \ldots
\label{eq:transform-open-closed}\\
x_{2k + 1} & = & \sum_{m = 1}^{\infty}
 Y_{2k + 1, 2m} \;y_m \;\;\;\;\; \;\;\;\;\;\; \; \; \;
 k = 0, 1, \ldots\nonumber
\end{eqnarray}
where
\begin{equation}
Y_{2m, 2k + 1} = Y_{2k + 1, 2m} =
(-1)^{m + k + 1} \frac{2m}{2k + 1}  X_{2m, 2k + 1}
= -\frac{8m}{\pi ((2k + 1)^2 -4m^2)} \,.
\end{equation}
is the transformation matrix between two bases on $\CH^-$, 
as follows from 
\begin{eqnarray}
\sin(2m\sigma) & =  & \sum_{k=0}^\infty Y_{2m,2k+1} \cos(2k+1)\sigma\qquad 0\leq \sigma<\pi \label{eq:sincos}\\
\cos(2k+1)\sigma & = & \sum_{m=1}^\infty Y_{2k+1,2m} \sin(2m\sigma) \qquad\qquad 0<\sigma<\pi \label{eq:cossin}
\end{eqnarray}
Here we have indicated the intervals of {\it pointwise} convergence. 
It follows from pointwise convergence of (\ref{eq:sincos}) at $\sigma =0$ that 
$x_{2k+1}=1$ is a zeromode of $Y$. Put differently, since   the vector $w$ is
annihilated by $X_{eo}$, it follows immediately that the vector with $
x_{2k + 1} =\tilde{w}_{2k + 1} = (2k + 1)w_{2k + 1} =1$ is annihilated by $Y$. 
 This vector is associated with the 
singular distribution $x (\sigma) = \sum_{k} \cos (2k + 1) \sigma $,
which behaves like the derivative of a delta function at $\sigma = 0$
when this point is identified with $\sigma = \pi$.

Using the transformation (\ref{eq:transform-open-closed}), we can
attempt to write a closed string state in terms of the open string
oscillators acting on the open string ground state.  For example,
consider the closed string vacuum $| 0 \rangle_c$ associated with the
closed string tachyon.  This state is described in closed string
coordinates by the functional (neglecting normalization factors)
\begin{equation}
| 0 \rangle_c \sim \exp\left({ -\sum_{m = 1}^{ \infty} m (y_m^2 +
  z_m^2) }\right) \,.
\label{eq:closed-vacuum-1}
\end{equation}
This expression might seem unfamiliar to some readers.  The standard
oscillators for the closed string (with $\alpha'=\half$) have
nonvanishing commutators $[\check{\alpha}_n, \check{\alpha}_m^\dagger] = \delta_{n,m}$, and
$[ \check{\tilde{\alpha}}_n,  \check{\tilde{\alpha}}_m^\dagger] = \delta_{n,m}$, $n,m=1,\dots,
\infty$. The oscillators
$\check{\alpha}_n^\dagger,  \check{\tilde{\alpha}}_n^\dagger$ create left and right-moving modes,
respectively. (The more conventional normalization is $\alpha_n =
\sqrt{n} \check{\alpha}_n, \tilde \alpha_n = \sqrt{n}
\check{\tilde{\alpha}}_n$.)
In standard closed string quantization the oscillators
are expressed in terms of $R$-valued coordinates $\xi_n$ and
momenta $\eta_n$. These   are canonically
conjugate: $[\eta_n , \xi_m] = -i \delta_{n,m}$.
The oscillators are then
$\check{\alpha}_n= {1\over \sqrt{2n}}(\eta_n - i n \xi_n)$
and can be written as
\begin{equation}
\check{\alpha}_n = -{i\over \sqrt{2n}} \bigl({\p \over \p \xi_n} + n \xi_n) =
{1 \over \sqrt{2n}} \bigl(n{\p \over \p \eta_n} + \eta_n)
\end{equation}
depending on the choice of polarization. Together with their 
right-moving analogues $\tilde\xi_n, \tilde \eta_n$ we have
\begin{equation} 
X(\sigma) = {1\over \sqrt{2}} \sum_{n=1}^\infty (\xi_n + \tilde \xi_n) \cos(2n\sigma) 
- {1\over \sqrt{2}} 
\sum_{n=1}^\infty {(\eta_n - \tilde \eta_n)\over n}  \sin(2n\sigma) 
\end{equation} 
while 
\begin{equation} 
P(\sigma) = {1\over \sqrt{2}} \sum_{n=1}^\infty (\eta_n + \tilde \eta_n) \cos(2n\sigma) 
+ {1\over \sqrt{2}} 
\sum_{n=1}^\infty n (\xi_n - \tilde \xi_n)  \sin(2n\sigma) 
\end{equation} 
Here $0 \leq \sigma \leq \pi$, and we have omitted the zeromodes. 
We  thus identify $z_n = \half (\xi_n + \tilde \xi_n)$ 
while $y_n = \half (\tilde \eta_n - \eta_n)/n$. 
Now, the standard closed string vacuum is usually written as 
\begin{equation} 
\prod_{n=1}^{\infty}( {\pi \over n})^{1/2} \exp[-\half n (\xi_n^2 + \tilde \xi_n^2) ] 
\end{equation} 
We may write $\xi_n^2 + \tilde \xi_n^2= \half ( \xi_n + \tilde \xi_n)^2 + 
\half (\xi_n - \tilde \xi_n)^2 $ and do  a Fourier transform on the variable 
$y_n$ conjugate to $ \half (\xi_n - \tilde \xi_n)$ 
 to produce the expression $\exp[-n (y_n^2 + z_n^2)]$.

Translating the closed string ground state (\ref{eq:closed-vacuum-1})
into open string coordinates using (\ref{eq:transform-open-closed}),
gives
\begin{equation}
\exp\left(-
 \sum_{m = 1}^{\infty}  m \;x_{2m}^2
-\sum_{k, l = 0, m = 1}^{ \infty} x_{2k + 1}\,
Y_{2k + 1, 2m} \; m\;  Y_{2m, 2l + 1}\, x_{2l + 1} \right)\,.
\label{eq:closed-vacuum-3}
\end{equation}
We would now like to rewrite this functional as a squeezed state
proportional to
\begin{equation}
| 0 \rangle_c \sim
\exp \left[ -\frac{1}{2}\sum_{n,m\geq 1} 
 a_m^{\dagger} K_{mn} a_n^{\dagger} \right]
| p = 0 \rangle\,.
\label{eq:closed-squeezed}
\end{equation}
Unfortunately, the summation over $m$ in the last term of
(\ref{eq:closed-vacuum-3}) is divergent, so a direct calculation of
the corresponding matrix $K$ is difficult.  Note, however, that the
sum is convergent whenever $\sum_{k} x_{2k + 1} = 0$ and $\sum_{k}
|x_{2k + 1} | < \infty$, as long as we first carry out the summations
over $k, l$.  This is because the terms of order $1/m$ in the matrix
element $Y_{2m, 2k + 1}$ are independent of $k$, and therefore cancel
when the $x_{2k + 1}$'s sum to $0$.  The divergence of the sum over $m$
when $\sum_{k} x_{2k + 1} \neq 0$ is a manifestation of the fact that
the functional (\ref{eq:closed-vacuum-3}) vanishes whenever $x_n$ has
a nonzero component in the direction $\tilde{\nu}$, which is exactly
what we expect for a squeezed state where $K$ has a -1 eigenvalue
associated with the vector $\tilde{\nu}$.  

While the preceding argument suggests that $K$ indeed has the claimed
eigenvalue, we will now carry out a second check on this claim
by the following indirect procedure: We will assume that the closed
string ground state can be written in the form
(\ref{eq:closed-squeezed}).  We then find the linear condition
satisfied by the open string creation and annihilation operators $a_n,
a^{\dagger}_n$ acting on the closed string vacuum.  This gives a
condition on the matrix $K$ which must be satisfied in order for
(\ref{eq:closed-vacuum-3}) to represent the closed string vacuum.  We
then show that this condition is compatible in a highly nontrivial
fashion with the condition that $\tilde{\nu}$ is an eigenvector of $K$
with eigenvalue -1.

In order to carry out this procedure we must first write the open
string creation and annihilation operators in terms of the closed
string coordinates $z_n, y_n$ and their derivatives.  We have
\begin{eqnarray}
a_{2k + 1} & = &  \frac{-i}{ \sqrt{2}}
\left[ \sqrt{2k + 1}\;Y_{2k + 1, 2m}\;y_{m}+ \frac{1}{ \sqrt{2k + 1}}
\;Y_{2m, 2k + 1}\; \frac{\partial}{ \partial y_{m}} \right]    \label{eq:a-yz}\\
a^{\dagger}_{2k + 1} & = &  \frac{-i}{ \sqrt{2}}
\left[ - \sqrt{2k + 1}\;Y_{2k + 1, 2m}\;y_{m}+ \frac{1}{ \sqrt{2k + 1}}
\;Y_{2m, 2k + 1}\; \frac{\partial}{ \partial y_{m}}   \right]\nonumber
\end{eqnarray}
For the squeezed state form of the closed string vacuum
(\ref{eq:closed-squeezed}) we have
\begin{equation}
\left(a_{2k + 1} + K_{2k + 1, 2l + 1} a^{\dagger}_{2l + 1}\right) | 0
\rangle_c = 0\,.
\label{eq:k-condition}
\end{equation}
Acting on the closed string vacuum functional
(\ref{eq:closed-vacuum-1}) using (\ref{eq:a-yz}) we find that for the
condition (\ref{eq:k-condition}) to be satisfied on the closed string
vacuum we must have, for all values of $m$,
\begin{eqnarray}
\lefteqn{\sum_{l}K_{2k + 1, 2l + 1} \left[ \sqrt{2l + 1} \;Y_{2l + 1, 2m}\;
+ \;Y_{2m, 2l + 1}\; \frac{2m}{ \sqrt{2l + 1}}  \right]
}\label{eq:k-determined} 
 \\ & = &  \left[ \sqrt{2k + 1} \;Y_{2k + 1, 2m}\;
- \;Y_{2m, 2k + 1}\; \frac{2m}{ \sqrt{2k + 1}}  \right]\,.\nonumber
\end{eqnarray}
Formally, this equation determines the matrix $K$ in the odd
sector to be 
\begin{equation}
K = (E^{-1}Y E - E Y E^{-1})(E^{-1}Y E + E Y E^{-1})^{-1} 
\end{equation}
while $K$ vanishes in the even sector.  We
are primarily interested in verifying that the matrix $K$ has a -1
eigenvalue, associated with the vector $\tilde{\nu}$.  We can check
this condition by acting on the left of (\ref{eq:k-determined}) with
the vector $\tilde{\nu}_k$ and summing over  $k$.  Since $\tilde{w}_k=
\sqrt{2k + 1} \; \tilde{\nu}_k$  is annihilated by $Y_{2k + 1, 2m}$, the
first term on the right hand side of (\ref{eq:k-determined}) cancels
and we are left with the condition
\begin{equation}
\sum_{k, l}\tilde{\nu}_kK_{kl} \left[ \sqrt{2l + 1} \;Y_{2l + 1, 2m}\;
+ \;Y_{2m, 2l + 1}\; \frac{2m}{ \sqrt{2l + 1}}  \right] =
-\sum_{k} \tilde{\nu}_k \;Y_{2m, 2k + 1}\; \frac{2m}{ \sqrt{2k + 1}}  \,.
\end{equation}
If $\tilde{\nu}$ is an eigenvector of $K$ with eigenvalue $-1$, we
again see that the first term on the left hand side cancels, and we
are left with a simple identity.  This shows that the defining
condition (\ref{eq:k-determined}) on the open string representation of
(the matter part of) the closed string vacuum is indeed compatible
with the desired role of $\tilde{\nu}$ as an eigenvector of $K$ with
eigenvalue -1 in a highly nontrivial fashion.  This gives a second
piece of evidence that the representation of the closed string vacuum
as a squeezed state (\ref{eq:closed-squeezed}) constructed from open
string raising operators is sensible, and that this squeezed state
obeys the relation
\begin{equation}
\left[\hat{x} (0) -\hat{x} (\pi) \right]
|0 \rangle_c = 0\,.
\label{eq:closed-vacuum-relation}
\end{equation}

Once we have identified the closed string vacuum as a state of the
form (\ref{eq:closed-squeezed}), it is straightforward to construct
all the excited closed string states in a similar fashion by acting
with closed string raising operators, reexpressed as linear
combinations of open string raising and lowering operators through
(\ref{eq:transform-open-closed}).  The resulting states give a
representation of the full (matter) closed string Hilbert space in
terms of states satisfying (\ref{eq:closed-relation}) which are
described by acting with open string raising operators on the open
string vacuum.  While these states are nonnormalizable states with
respect to the open string Hilbert space, they are no worse behaved
than the sliver states we have discussed in previous sections, the
Dirac position and momentum basis states in finite-dimensional quantum
mechanics, or the boundary states used frequently to discuss D-branes
in closed string theory.

There has been much debate about whether closed strings can be seen in
a natural fashion in the classically stable vacuum of Witten's cubic
string field theory which arises after the open string tachyon
condenses.  While it is known that closed string poles can be seen in
the one-loop two point function of the open string field theory
\cite{closed-poles}, it would be more satisfying if these closed
string states could be explicitly constructed as asymptotic states in
the open string field theory.  Some approaches to describing closed
string states in terms of open string field theory were developed in
\cite{Strominger,Yi-closed,bhy,ghy,Sen-closed,Gerasimov-Shatashvili,Shatashvili-closed}.
The discussion we have just given indicates the form that these closed
string states should take.  While these states are presumably
nonnormalizable poles of the one-loop two-point function, a sequence
of finite matrix approximations to the full two-point function given
by successive level truncations should give a sequence of approximate
poles approaching the desired closed string state.  (This situation is
analogous to considering a sequence of matrix approximations to the
operator $x$ acting on $L^2 (R)$ using a finite set of harmonic
oscillator eigenstates--while the Dirac delta function $\delta (x)$ is
not a normalizable state in the Hilbert space, the sequence of lowest
eigenvalues of this sequence of matrices approaches 0, and the
associated eigenvectors give successive approximations to the squeezed
state representation of the delta function, $e^{-a^{\dagger}
a^{\dagger}/2} | 0 \rangle$).  It would be very desirable, although
technically challenging, to check this level-truncated string field
theory calculation explicitly.

If it is indeed possible to  sensibly define all closed string states
in the language of open string oscillators, it will be strong evidence
that the open string field theory in the tachyonic vacuum in fact
contains all information about closed string diagrams needed to
construct a consistent closed string field theory.  Indeed, it was
shown by Giddings, Martinec, and Witten \cite{gmw} that the set of
open string diagrams produced by the cubic open string field theory
precisely covers the moduli space of Riemann surfaces of arbitrary
genus with a nonzero number of boundary components.  By expressing the
closed string states in the open string language, we are essentially
contracting these boundary components to pointlike punctures, so that
the moduli space of all closed string diagrams is naturally covered.
Issues which may be related to this picture are discussed in
\cite{rsz-forthcoming}

Finally, the present discussion has some similarity with the theory of boundary
states in conformal field theory. Here one can view open string
oscillators as the quantization of the constrained subspace of the
closed string phase space defined by the second class constraints
$(\alpha_n - \tilde \alpha_n) =0 , \forall n\in Z$, (NN conditions) or
$(\alpha_n + \tilde \alpha_n) =0 , \forall n\in Z$, (DD conditions).
The boundary state $\vert B\rangle\rangle = \exp[\pm
\sum_{n=1}^\infty{1\over n} \alpha_{-n} \tilde \alpha_{-n} ] \vert 0
\rangle_{closed}$ is the squeezed state satisfying $(\alpha_{n} \mp
\tilde \alpha_{-n}) \vert B \rangle\rangle =0$.  In terms of the open
string oscillators it is the singular squeezed state $\exp[\pm \half
\sum_{n=1}^\infty a_n^\dagger a_n^\dagger ] \vert 0 \rangle_{open}$
and the analog of the K-matrix has eigenvalue $\pm 1$.  It might be
interesting to pursue further the analogies with boundary states.

\section{Discussion}
\label{sec:discussion}


It should be clear to the reader that the considerations of this paper
can and should be generalized in several ways. First, 
we expect that it is a fairly straightforward exercise 
to include the $B$ field in the above analysis, so we can 
take the Seiberg-Witten limit to recover the 
relation between D-branes and noncommutative geometry 
\cite{Gross-Nekrasov,harvey-komaba,douglas-nekrasov}. This should make 
contact with some of the considerations of \cite{bars}.  To be
slightly more explicit, it should be possible, starting with the
3-string vertex in a general background metric and $B$ field as
calculated in \cite{sugino,kawano} to calculate the general analogue
of the sliver state by taking the limit of an infinite product of
ground states $| 0 \rangle$.  It should then be possible to take the
$t \rightarrow \infty$ limit in which the star product factorizes into
a space-time Moyal product algebra and the usual $p = 0$ star product
algebra, as in \cite{wittencomment}.  In this limit the sliver should
become the solitonic projection operator of \cite{gms}.  This can be
seen directly by noting that the noncommutative soliton projector
arises as the limit of an infinite Moyal product of a space-time
Gaussian with itself.  If we then take the $B$ field to vanish we
return to the singular projection operator found in Section 3.  It is
interesting to note that in the $t \rightarrow \infty$ limit the
singular structure of the projection operator should be smoothed out by the
presence of the $B$ field.  Since the singularity arises from the -1
eigenvalue of $S'$ for any $t$, this singularity must also be smoothed
by the introduction of a $B$ field at arbitrary $t$, suggesting that
there is a nontrivial nonsingular analogue of the D-instanton
projection operator in nonzero $B$ field.  Note, however, that the
string splitting behavior we have found for the sliver in the
longitudinal directions (all directions for the D25-brane) will not be
affected by the introduction of a $B$ field as it arises purely from
the odd sector of $V$, which is not affected by a $B$ field.  

Another interesting question arising from this work
is how one describes the zero-slope limit of a configuration of
multiple D-instanton slivers.   The most natural way to make 
multiple D-branes is to consider two branes at 
relative separation $y$ and let these approach one another. 
The translated sliver is $\exp[ y {\p \over \p x_0}]\vert \Xi_0\rangle $. 
The expectation value 
\begin{equation} 
{\langle \Xi_0 \vert \exp[ y {\p \over \p x_0}] \vert \Xi_0\rangle
\over
\langle \Xi_0 \vert \Xi_0\rangle}
=   \exp\bigl[\half y^2 - \langle 0 \vert {1\over 1+S'} \vert 0 \rangle y^2 \bigr]
\label{eq:ratio} 
\end{equation}
is ill-defined because $S'$ has an eigenvalue $-1$. One should
probably not conclude from this that the sliver and its translate are
orthogonal. Rather one should probably introduce a regularization. We
believe (but have not proved) that if we regularize $S'_{nm} \to
q^{n+m} S'_{nm} $ then the zero eigenvalue of $1+S'$ is lifted to
$\epsilon'(\epsilon,q)$. We believe that $\epsilon'(\epsilon,q)$ goes
to $0$ if $q\to 1$, for any $\epsilon$. Moreover, for $\epsilon\to 0$
the null eigenvector of $1+S'$ becomes parallel with $\vert 0
\rangle$. Therefore, in the limit $y \to 0, \epsilon\to 0 , q\to 1$
(\ref{eq:ratio}) becomes $\exp[-y^2/\epsilon'(\epsilon,q)]$ and hence
we can take a scaling limit such that two infinitesimally separated
slivers are not orthogonal.  The situation is now completely parallel
to that of multiple noncommutative solitons
\cite{Gross-Nekrasov,harvey-komaba,douglas-nekrasov}, and based on our
experience 
with these we should be able to recover higher rank projectors. But
many details of this proposal remain to be filled in.

One of the main applications of sliver states in previous work has
been the construction of D$p$-brane-like solutions in the RSZ vacuum
string field theory model \cite{rsz-2,rsz-3,Gross-Taylor-I}.  The
results described here on the singular structure of sliver states may
help to understand some aspects of how the RSZ model fits into the
framework given by Witten's original cubic string field theory.  All
of the sliver states we have discussed here have norms which are
formally either vanishing or infinite.  The problems with the norms of
these states arise from the $\pm 1$ eigenvalue of the matrices $S, S'$
controlling the squeezed state representations of the sliver.
Numerical evidence on the stable vacuum of Witten's string field
theory \cite{Moeller-Taylor} suggests that this state is better
behaved than the sliver states discussed in this paper.  Defining a
norm on the ghost sector using the operator $c_0$, we find that the
contribution at level $L$ to the norm of the stable vacuum state
decreases faster than $1/L$, even when the absolute value of each
contributing term is added.  This suggests, but does not prove, that
the stable vacuum state of Witten's cubic string field theory in
Feynman-Siegel gauge is normalizable with respect to an appropriate
inner product.  The fact that the sliver states are not normalizable states in
the Hilbert space suggests that the RSZ model is a somewhat singular
limit of Witten's cubic string field theory around the stable vacuum
(assuming that the sliver states are relevant for describing D$p$-branes in
the RSZ vacuum string field theory).  Indeed, the results of
\cite{Gross-Taylor-II} on the ghost sector of the RSZ model, which
show that the action vanishes for any solution lying in the Hilbert
space, give another indication of this singular nature of the RSZ
model.  It is not surprising, since the RSZ model completely decouples
the matter and ghost sectors of the theory, that the separate
anomalies which arise in the two sectors cause this model to be
singular in some features.  The interesting question is whether these
singularities can be dealt with in a useful way to achieve new insight
into the theory.  It may be that the role of the ghosts and the more
complicated BRST operator of the Witten theory is precisely to
regulate these singularities in a consistent and well-defined way so
that the D-brane states, which naively seem singular from the point of
view of the matter theory, become well-behaved states in the open
string Hilbert space.

A fundamental question about open string field theory, which has
troubled workers in this area since the early days of the subject, is
precisely what set of states should be allowed in the string field
theory star algebra.  As discussed in section 4.3, the matter star
product is not closed on the open string matter Hilbert space.  In
this paper we have described a number of interesting states which have
a similar behavior to Dirac's position and momentum basis states for
quantum mechanics.  These states do not lie in the Hilbert space, but
are ``very close'' to lying in the Hilbert space, in the sense that
they can be described as suitable limits of well-behaved states.  We
have shown that this category of states which are ``almost'' in the
open string Hilbert space includes the matter sliver states
corresponding to D$p$-branes of all dimensions, as well as the
spectrum of closed string states described in terms of open string
oscillators.  One conservative approach would be to mandate that no
states outside the Hilbert space are allowed at all.  In this case we
must either reject the sliver and closed string states from the
theory, or hope that the inclusion of ghosts in Witten's theory serves
to regulate the singularities associated with these states.  Another
possibility, however, is that certain states outside the conventional
open string Hilbert space must be included in the star algebra.
Indeed, several arguments suggest that such states {\it must} be
included for the theory to be consistent and to contain all the
desired physics.  The first of these arguments arises simply from the
natural suggestion that the open string star algebra be closed, and
the observation that this algebra does not close on the Hilbert space.
Another argument arises from considering string states on a D$p$-brane
which has been translated in a transverse direction.  These string
states satisfy Dirichlet boundary conditions with different values
than those associated with the original D$p$-brane, and thus lie
outside the Hilbert space of DD strings on the original D-brane.  We
would clearly like to be able to describe translation of D-branes in
open string field theory.  If this physical requirement on the theory
is to be met it, is therefore necessary to include states in the star
algebra which are not in the original open string Hilbert space of the
conformal field theory used to define the Witten cubic SFT.  Some
progress in this direction was made in
\cite{Sen-Zwiebach-marginal,tt}, where an attempt was made to
explicitly construct (the T-dual of) a translated D$p$-brane in a
level truncation of Witten's cubic SFT in Feynman-Siegel gauge.  It
was found that the solution associated with the translated D$p$-brane
encounters a singularity at a distance on the order of the string
length.  This is where we would expect a state to leave the
normalizable string Hilbert space, so this result is perhaps not
surprising.  These arguments, however, suggest that it is necessary to
include certain states somewhat outside the original normalizable open
string Hilbert space to make Witten's cubic string field theory
consistent.  If such states are added, it seems clear from the
argument of the previous section that we should naturally expect
closed string states to be incorporated into the theory.  It may be
that the most natural way to understand the appearance of all these
singular states is by using the level-truncated theory.  The regulator
imposed by level truncation has the effect of naturally rendering
finite the divergences associated with states like the sliver state
and closed string states, so that we would hope to see some indication
of these states in a finite approximation to the full open string
field theory.

\section*{Acknowledgements}

We would like to thank M.\ Douglas, I.\ Ellwood, D.\ Gross,
H.\ Liu, J.\ Michelson,
M.\ Schnabl, A.\ Sen, I.\ M.\ Singer, and B.\ Zwiebach for
discussions.  We would like to thank the ITP, Santa Barbara, and the
ITP workshops on M-theory and mathematical physics for support and
hospitality during the progress of this work.  WT would like to thank
the high energy theory group at Rutgers University for hospitality
during the latter stages of this work.  The work of GM was supported
by DOE grant DE-FG02-96ER40949.  The work of WT was supported in part
by the A.\ P.\ Sloan Foundation and in part by the DOE through
contract \#DE-FC02-94ER40818.

\normalsize

\bibliographystyle{plain}

\end{document}